\documentclass{amsart}
\usepackage{amssymb}
\usepackage{amsfonts}
\usepackage{amsmath}
\usepackage{graphicx}
\usepackage{epstopdf}
\usepackage[backref]{hyperref}

\begin{document}

	\begin{center}
		{\Large \bf On partition functions of refined Chern-Simons \\ theories on $S^3$ \\
			\vspace*{1 cm}
			
			{\large  M.Y. Avetisyan and R.L.Mkrtchyan
			}
			\vspace*{0.2 cm}
			
			{\small\it Yerevan Physics Institute, 2 Alikhanian Br. Str., 0036 Yerevan, Armenia}
			
		}
		
	\end{center}\vspace{2cm}

{\small  {\bf Abstract.} 

We present a new expression for the partition function of refined Chern-Simons theory on $S^3$ with arbitrary gauge group,
 which is explicitly equal to $1$, when the coupling constant is zero.
Using this form of partition function we show that the previously known Krefl-Schwarz representation of partition function of refined Chern-Simons theory on $S^3$ can
 be generalized to all simply-laced algebras.

 For all non-simply-laced gauge algebras we derive similar representations of that partition function, which makes it
possible to transform it into a product of multiple sine functions aiming at the further establishment of a duality with refined topological strings.


\section{Introduction}

Partition function of Chern-Simons (CS) theory on a three-dimensional sphere $S^3$, first calculated in \cite{W1} (see below (\ref{Zk})), is represented
in the universal form in \cite{MV,M13}, which means that alternative to the pure Lie algebra data - roots, invariant scalar product, etc., it is now
expressed in terms of the so-called Vogel's universal parameters $\alpha, \beta, \gamma$ \cite{V,V0}. 
These universal parameters are the homogeneous coordinates of a projective plane, which is called Vogel's plane in this context, so that special points in this plane
 correspond to all the simple Lie algebras (see Vogel's table \ref{tab:1}).
 The projective nature of universal parameters, i.e. their rescaling property, is the reflection of the possibility of rescaling of the invariant scalar product 
  in the Lie algebra \cite{V0,LM}
 
The advantage of this representation is that it is very convenient for further transformation of the abovementioned partition function into the Gopakumar-Vafa partition function of topological strings, 
as shown in \cite{M13,SGV} for the Chern-Simons theory with classical groups. In the recent work \cite{M21} this transformation has been
 extended to the CS with the exceptional groups, meaning that the partition function of CS on $S^3$ with an exceptional gauge group 
 has been presented in the form of a partition function of specific refined topological string. This should be considered as a step towards the establishment of the duality of the corresponding theories.
 The fact that all exceptional algebras (actually all algebras in $E_8$ row of the Freudenthal magic square) belong to a line in Vogel's plane - the so-called Deligne's line, is exploited in that work. 
Deligne \cite{Del96} suggested that all points on that line make up a so-called series of Lie algebras, which was partially confirmed in \cite{Cohen}.

The main features of representation of partition function discovered in \cite{MV,M13} has been extended to include the partition function of the refined Chern-Simons theory $S^3$ sphere
\footnote{we shall omit mentioning $S^3$ from now on, 
since we don't consider theories on other manifolds in this paper} for $A_n$ and $D_n$ algebras in \cite{KS}. It has also been shown to be very convenient for derivation of the partition
 function of dual refined topological strings in \cite{KM}. 
In the same work the non-perturbative corrections to the partition function of topological strings, derived from the universal Chern-Simons partition function \cite{M13} 
(with $A_n$ gauge algebra), has been shown to coincide with those, derived in \cite{HMMO13,H15} directly in topological string theory,
 thus extending the CS/topological strings duality to the non-perturbative domain. 

The natural development of these investigations would be the extension of the universal-type representation of the refined CS theories with $A_n$ and $D_n$ algebras to the remaining 
algebras: the simply-laced $E_n$ and the non-simply-laced classical ($B_n, C_n$) and exceptional ($F_4, G_2$) algebras, with the final aim of setting up a connection of the corresponding 
refined CS theories with some (refined) topological strings. 

Present paper embodies the first step in this direction: we present, for the first time, the universal-type representations of partition function of the refined CS theory  
with each of the remaining gauge groups. 
The transformation of these partition functions into those of specific refined topological strings is under investigation \cite{AM-21}

In Section \ref{partfunc} we present the new representation of the partition function of refined CS theory for {\it all} simple Lie algebras. 
It is based on a new Lie-algebraic identity for the determinant of the symmetrized Cartan matrix (refined version of that in \cite{KP}), and generalizes a feature of the non-refined 
theory, exploited in \cite{M13} earlier, which states that partition function is equal to $1$ when the coupling of CS is $0$. 

Then, in section \ref{sect:univ} we represent this partition function in the "universal" form, which means that instead of the roots and other standard characteristics of a gauge algebra it
now writes in terms of the Vogel's parameters. Simultaneously, the range of the refinement parameter is extended to include non-integer values, too. 

We discuss the prospects of further transformation of this partition function into partition functions of refined topological strings in Conclusion.

\section{The partition function of refined Chern-Simons theory on $S^3$} \label{partfunc}

The partition function of Chern-Simons theory on $S^3$ sphere was given in Witten's seminal paper \cite{W1} as the $S_{00}$ element of $S$ matrix of modular transformations. 
For an arbitrary gauge group it is (see, e.g. \cite{DiF,MV})

 \begin{eqnarray} \label{Zk}
 	Z(k)= Vol(Q^{\vee})^{-1} (k+h^{\vee})^{-\frac{r}{2}} \prod_{\alpha_+} 2\sin \pi \frac{(\alpha,\rho)}{k+h^{\vee}}
 \end{eqnarray}
Here the so called minimal normalization of invariant scalar product $(,)$ in the root space is used, which implies that the square of the long roots equals $2$. 
Other notations are: $Vol(Q^{\vee})$ is the volume of the fundamental domain of the coroot lattice $Q^\vee$, integer $k$ is  the Chern-Simons coupling constant, 
$h^\vee$ is the dual Coxeter number of the algebra, $r$ is the rank of the algebra,  
the product is taken over all positive roots $\alpha_+$. 

 $Vol(Q^{\vee})$ is equal to the square root of the  determinant of the matrix of scalar products of simple coroots,  accordingly for the simply laced algebras, in the minimal normalization, it is equal to the square root of the determinant of Cartan matrix:   

\begin{eqnarray}
	Vol(Q^{\vee})= (\det (\alpha_i^\vee,\alpha_j^\vee))^{1/2} \\
	\alpha_i^\vee=\alpha_i \frac{2}{(\alpha_i,\alpha_i)}, \,\,
	i=1,...,r
\end{eqnarray}

The same formula for partition function, rewritten in the arbitrary normalization of the scalar product \cite{MV}, is

\begin{eqnarray} \label{Zkap}
	Z(\kappa)= Vol(Q^{\vee})^{-1} (\delta)^{-\frac{r}{2}} \prod_{\alpha_+} 2\sin \pi \frac{(\alpha,\rho)}{\delta}
\end{eqnarray}

where $k$ is now replaced by $\kappa$, $h^\vee$ by $t$, and  $\delta=\kappa+t$. 
In this form the r.h.s. is invariant w.r.t. the simultaneous rescaling of the scalar product, $\kappa$, and $t$ (and hence $\delta$). In minimal normalization they accept their usual values in (\ref{Zk}).

In \cite{M13} it was noticed, that from this formula for partition function one can derive interesting closed expression for $Vol(Q^{\vee})$, which agrees with that in the Kac-Peterson's paper \cite{KP}, (see eq. (4.32.2)), provided

 \begin{eqnarray} \label{Z0}
  Z(0)=1
  \end{eqnarray}

This equality is completely natural from the physical point of view. Indeed, the Chern-Simons theory is based on the
unitary integrable representations of affine Kac-Moody algebras. At a given $k$ there is a finite number of such representations, and at $k=0$ there is not any non-trivial one. 

So, from (\ref{Zkap}) and (\ref{Z0}) we have

\begin{eqnarray} \label{voly=1}
	 Vol(Q^{\vee})= t^{-\frac{r}{2}} \prod_{\alpha_+} 2\sin \pi \frac{(\alpha,\rho)}{t}
\end{eqnarray}
 which, as mentioned, agrees with \cite{KP}. Below we generalize this equation by inclusion of a refinement parameter.

Generalization of Chern-Simons theory to the  refined Chern-Simons theory is given in \cite{AS11,AS12a,AS12}. 
It is based on the Macdonald's deformation of e.g. the Shur polynomials, and other  "deformed" formulae, given in \cite{Mac1,Mac2,Mac3}. 
In a nutshell, the Macdonald's deformation yields deformed $S$ and $T$ matrices of the modular transformations, and since these matrices define all observables in Chern-Simons theory,
one can  naturally consider the "deformed" or refined versions of all observables, i.e. the link/manifold invariants.

Particularly, the partition function of refined Chern-Simons theory on $S^3$ is given \cite{AS11} by the $S_{00}$ element of the refined $S$-matrix.
In \cite{AS11} an orthogonal, instead of an orthonormal basis is used sometimes. We shall use the orthonormal one only (as in \cite{KS}), so  there is no difference between e.g. $S_{00}$ and $S_0^0$.

We suggest the following expression for $S_{00}$ for the refined CS theory:

 \begin{eqnarray}\label{refCS}
	Z(\kappa,y)= Vol(Q^{\vee})^{-1} \delta^{-\frac{r}{2}} \prod_{m=0}^{y-1} \prod_{\alpha_+} 2\sin \pi \frac{y(\alpha,\rho)-m (\alpha,\alpha)/2}{\delta}
\end{eqnarray}

We assume that now $\delta=\kappa+y t$, $y$ is the refinement parameter, which we consider to be a positive integer at this stage. 

Although we could not find the $Z(\kappa,y)$ in this exactly form in literature, however, the expression (\ref{refCS}) complies with the known formulae in different limits, 
e.g. at $y=1$ it yields the corresponding formula for the non-refined case (\ref{Zkap}). It also coincides with the corresponding formulae for refined CS theory 
in \cite{AS11,AS12,KS} for $A_n, D_n$ algebras. 
The coefficient $(\alpha,\alpha)/2$ in front of the summation parameter $m$ coincides with that in the constant term formulae in \cite{Cher1,Cher2}.
 Actually for not-simply-laced algebras one can introduce two refinement parameters, one for each length of roots, see e.g. \cite{Cher1,Cher2}. 
 However, we did not try to introduce a second parameter (and also are not aware of the physical interpretation of it), so below we consider them to be coinciding,
  so that we always have one refinement parameter.

The latter expression of the partition function is supported by the key feature of (\ref{refCS}): at $\kappa=0$ the equality $Z(0,y)=1$ holds, 
which is ensured by the following generalization of the formula (\ref{voly=1}) for the same object  $Vol(Q^{\vee})$:

\begin{eqnarray} \label{vol}
	 Vol(Q^{\vee})=  (ty)^{-\frac{r}{2}}  \prod_{m=0}^{y-1} \prod_{\alpha_+} 2\sin \pi \frac{y(\alpha,\rho)-m (\alpha,\alpha)/2}{ty}
\end{eqnarray}

For $A_{n}$ algebras this equality can be easily proved with the use of the following well-known identity, valid at an arbitrary positive integer $N$:
\begin{eqnarray}
	N=\prod_{k=1}^{N-1}2 \sin{\pi\frac{k}{N}}
\end{eqnarray}
Similarly it can be checked for all the remaining root systems, too. 

Next, with (\ref{vol}) taken into account, we obtain the following
expression of the partition function:

\begin{eqnarray}
	Z(\kappa,y)= \left(\frac{ty}{\delta}\right)^{\frac{r}{2}} \prod_{m=0}^{y-1} \prod_{\alpha_+} \frac{\sin \pi \frac{y(\alpha,\rho)-m (\alpha,\alpha)/2}{\delta}}{\sin \pi \frac{y(\alpha,\rho)-m (\alpha,\alpha)/2}{ty}}
\end{eqnarray}

which explicitly satisfies $Z(0,y)=1$, since $\delta=t y$ at $\kappa=0$.

\section{Integral representation of partition function for the refined CS theories}

For this purpose we apply the transformation, introduced in \cite{M13}. We transform each of the sines into a pair of Gamma-functions by the following 
well-known identity

\begin{eqnarray}
	\frac{\sin \pi z}{\pi z} = \frac{1}{\Gamma(1+z) \Gamma(1-z)}
\end{eqnarray}

and make use of the integral representation of (the logarythm of) the $\Gamma$ function:

\begin{eqnarray}
	\ln\Gamma(1+z)=\int_{0}^{\infty}dx \frac{e^{-zx}+z(1-e^{-x})-1}{x(e^{x}-1)}
\end{eqnarray}

Let us rewrite the partition function in the following form:

\begin{eqnarray}
Z(\kappa,y)=\left(\frac{ty}{\delta}\right)^{y\frac{dim-r}{2}+\frac{r}{2}}	\prod_{m=0}^{y-1} \prod_{\alpha_+} \frac{\sin \pi \frac{y(\alpha,\rho)-m (\alpha,\alpha)/2}{\delta}}{\pi \frac{y(\alpha,\rho)-m (\alpha,\alpha)/2}{\delta}} \times \\
 \prod_{m=0}^{y-1} \prod_{\alpha_+}\frac{\pi \frac{y(\alpha,\rho)-m (\alpha,\alpha)/2}{ty}}{\sin \pi \frac{y(\alpha,\rho)-m (\alpha,\alpha)/2}{ty}} \equiv\\
 \left(\frac{ty}{\delta}\right)^{y\frac{dim-r}{2}+\frac{r}{2}} Z_1 Z_2
\end{eqnarray}

and apply the abovementioned transformation to the first couple of products of sines (then similarly to the second couple of products):

\begin{eqnarray}
\ln Z_1=	\ln \left( \prod_{m=0}^{y-1} \prod_{\alpha_+} \frac{\sin \pi \frac{y(\alpha,\rho)-m (\alpha,\alpha)/2}{\delta}}{\pi \frac{y(\alpha,\rho)-m (\alpha,\alpha)/2}{\delta}} \right)= \\
	-\int_{0}^{\infty} \frac{dx}{x(e^{x}-1)} \sum_{m=0}^{y-1}\sum_{\alpha_+}\left( e^{x\frac{y(\alpha,\rho) -m (\alpha,\alpha)/2}{\delta}}+e^{-x\frac{y(\alpha,\rho) -m (\alpha,\alpha)/2}{\delta}}-2\right)
\end{eqnarray}

Let us introduce the following function for any simple Lie algebra $X$ of rank $r$:

\begin{eqnarray}\label{FX2}
		F_X(x,y)=    r+	\sum_{m=0}^{y-1}	\sum_{\alpha_{+}} \left( e^{x(y(\alpha,\rho) -m (\alpha,\alpha)/2)}+e^{-x(y(\alpha,\rho) -m (\alpha,\alpha)/2)}\right)
\end{eqnarray}

Then 

\begin{eqnarray}
	\sum_{m=0}^{y-1}	\sum_{\alpha_+}\left( e^{x(y(\alpha,\rho) -m (\alpha,\alpha)/2)}+e^{-x(y(\alpha,\rho) -m (\alpha,\alpha)/2)}-2\right)= \\
	F_X(x,y) - r-y(dim-r)
\end{eqnarray}
and  $\ln Z_1$ becomes

\begin{eqnarray}
	\ln Z_1= -\int_{0}^{\infty} \frac{dx}{x(e^{x}-1)} \left( F_X\left( \frac{x}{\delta},y\right) - r-y(dim-r)\right)
\end{eqnarray}

Similar transformation applies to $\ln Z_2$:

\begin{eqnarray}
	\ln Z_2= \int_{0}^{\infty} \frac{dx}{x(e^{x}-1)} \left( F_X\left( \frac{x}{ty},y\right) - r-y(dim-r)\right)
\end{eqnarray}

and $\ln Z$ takes the form

\begin{eqnarray} \label{Zintg}
	\ln Z= \\
	\frac{1}{2} (y(dim-r)+r)  \ln \left(\frac{ty}{\delta}\right)  +\int_{0}^{\infty} \frac{dx}{x(e^{x}-1)} \left( F_X\left( \frac{x}{t y},y\right) - F_X\left( \frac{x}{\delta},y\right) \right)
\end{eqnarray}

Finally, one can further transform this formula into expression, similar to that which was derived in \cite{KM} for unrefined theories. 

Let us make the $x\rightarrow ty x/\delta$ rescaling  in $\ln Z_2$, so that
\begin{eqnarray}
	\ln Z_2=\int_0^\infty \frac{dx}{x(e^{ty x/\delta}-1)} \left(F_X\left(\frac{x}{\delta},y\right)- r-y(dim-r)\right)\,.
\end{eqnarray}

Using the relation
\begin{eqnarray} \label{cotmcotId}
	\frac{1}{e^{b x}-1}-\frac{1}{e^{a x} -1} = \frac{e^{a x}-e^{b x}}{(e^{a x}-1)(e^{b x }-1)}=\frac{\sinh\left(\frac{x(a-b)}{2}\right)}{2\sinh\left(\frac{x a}{2}\right)\sinh\left(\frac{x b}{2}\right)}\,,
\end{eqnarray}
and making use of that the combined integrand is even under $x\rightarrow -x$, 
we can write $\ln Z$ as 
\begin{eqnarray} \label{FVpreFinal}
	\ln Z=\frac{ r+y(dim-r)}{2} \log \left(ty/\delta\right)- \\
	\frac{1}{4}\int_{R_+} \frac{dx}{x} \frac{\sinh\left(x(ty-\delta)\right)}{\sinh\left(x ty \right)\sinh\left(x \delta\right)} \left(F_X(2x,y)- r-y(dim-r)\right)\,,
\end{eqnarray}

where the integration range passes the origin by an infinitesimal semi-circle in the upper (or lower) half of the complex plane. We denote the 
corresponding contour $R_+ $. We also took $x\rightarrow 2x \delta$. 

Due to the following identity
\begin{eqnarray}
	\frac{1}{4}\int_{R_+} \frac{dx}{x} \frac{\sinh\left(x(t-\delta)\right)}{\sinh\left(x t\right)\sinh\left(x \delta\right)}=-\frac{1}{2}\log\left(\frac{t}{\delta}\right)\,,
\end{eqnarray}
proved in \cite{KS} the integral of the $r+y(dim-r)$ term in fact cancels against the $\log$ term in \ref{FVpreFinal}, so that we obtain the final expression:

\begin{eqnarray}\label{FintRep}
	\ln Z=-\frac{1}{4}\int_{R_+} \frac{dx}{x} \frac{\sinh\left(x(ty-\delta)\right)}{\sinh\left( x ty \right)\sinh\left(x \delta\right)} F_X(2x,y)
\end{eqnarray}

With a corresponding representation of $F_X(x,y)$ functions as a ratio of polynomials over $q=\exp x$, which is shown below in section 6, the latter 
expression can be transformed into a product of multiple sine functions (see, e.g. \cite{SGV,KM}), which then hopefully will make the further correspondence of it with refined topological strings possible.

\section{Partition function of refined CS for simply-laced algebras}

In the non-refined case, i.e. at $y=1$ (when the sum over $m$ disappears) the partition function rewrites in terms of the Vogel's universal parameters.
The corresponding $F_X(x,1)$ coincides with quantum dimension of the adjoint representation, which is the character
 $\chi_{ad}(x\rho)$, restricted to the $x\rho$ line, collinear with the Weyl vector $\rho$: 
\begin{eqnarray}
	F_X(x,1)=r+\sum_{\alpha_+}\left( e^{x(\alpha,\rho) }+e^{-x(\alpha,\rho)}\right)=	\chi_{ad}(x\rho)
\end{eqnarray}

So $F_X(x,y)$ can be called refined quantum dimension.

Quantum dimension of the adjoint representation has been represented in the universal form in \cite{W3,MV}: 

\begin{eqnarray}\label{dim}
		\chi_{ad}(x\rho) \equiv
		f(x)		&=& \frac{\sinh(x\frac{\alpha-2t}{4})}{\sinh(x\frac{\alpha}{4})}\frac{\sinh(x\frac{\beta-2t}{4})}{\sinh(x\frac{\beta}{4})}\frac{\sinh(x\frac{\gamma-2t}{4})}{\sinh(x\frac{\gamma}{4})}
	\end{eqnarray}

Note that the notation $\alpha$ is used either for the root(s) of an algebra and for one of the Vogel's parameters. Since these objects are very different, hopefully no interpretation problem will appear.

Finally, the partition function in the non-refined case takes the following universal form

\begin{eqnarray}
Z(\kappa)=	Z(\kappa,1)= \left(\frac{t}{\delta}\right)^{\frac{dim}{2}} exp \left( -\int_{0}^{\infty} \frac{dx}{x(e^{x}-1)} \left( f\left(\frac{x}{\delta}\right)-f\left(\frac{x}{t}\right)  \right)      \right)
\end{eqnarray}
first given in \cite{M13}

In the refined case there is not a similar universal answer for the double sum over $m$ and $\alpha_+$, however, for $A_n$ and $D_n$ algebras 
Krefl and Schwarz \cite{KS} have made a statement, equivalent to 

\begin{eqnarray} \label{fxd}
\sum_{m=0}^{y-1}	\sum_{\alpha_+}\left( e^{x(y(\alpha,\rho) -m (\alpha,\alpha)/2)}+e^{-x(y(\alpha,\rho) -m (\alpha,\alpha)/2)}-2\right)=  \\ f\left(x,y\right)-dim(y) 
\end{eqnarray}

with
\begin{eqnarray}
	f(x,y)= \frac{\sinh(x\frac{\alpha-2ty}{4})}{\sinh(x\frac{\alpha}{4})}\frac{\sinh(xy\frac{\beta-2t}{4})}{\sinh(xy\frac{\beta}{4})}\frac{\sinh(xy\frac{\gamma-2t}{4})}{\sinh(xy\frac{\gamma}{4})},  \\
	dim(y)= \lim_{x\rightarrow 0} f(x,y) = y \,\, dim-(y-1)\frac{(\beta-2t)(\gamma-2t)}{\beta \gamma} \\
	f(x,1)=f(x)
\end{eqnarray}
where it is assumed that $\alpha$ is the only negative parameter (equal to $-2$ in the minimal normalization of the scalar product).

The  $dim(y)$ can be further transformed. Indeed, consider the dimension formula of simple Lie algebras:

\begin{eqnarray}
	dim=\frac{(\alpha-2t)(\beta-2t)(\gamma-2t)}{\alpha \beta \gamma} =\frac{\alpha-2t}{\alpha}  \frac{(\beta-2t)(\gamma-2t)}{ \beta \gamma}
\end{eqnarray}
In the last expression both fractions are independent of the normalization. In the minimal normalization the
 first fraction is equal to $1+h^\vee$ (where $h^\vee$ is the dual Coxeter number) so we conclude that the second one is the rank of the algebra

\begin{eqnarray}
	 \frac{(\beta-2t)(\gamma-2t)}{ \beta \gamma} =r
\end{eqnarray}

since the following relation holds for all simply-laced algebras:

\begin{eqnarray}
	dim = (1+h^\vee) r
\end{eqnarray}

Finally, we have 
\begin{eqnarray}
	dim(y)=y(dim-r)+r
\end{eqnarray}

With this relation we see that (\ref{fxd}) is equivalent to 

\begin{eqnarray} \label{F=f}
	F_X(x,y)=f(x,y)
\end{eqnarray}

Then, with the use of (\ref{fxd}), the partition function (\ref{Zintg}) becomes:

\begin{eqnarray} \label{zky}
	Z(\kappa,y)= \left(\frac{ty}{\delta}\right)^{y\frac{dim-r}{2}+\frac{r}{2}} exp \left( -\int_{0}^{\infty} \frac{dx}{x(e^{x}-1)} \left( f\left(\frac{x}{\delta},y\right)-f\left(\frac{x}{ty},y\right)  \right)      \right)
\end{eqnarray}

As mentioned, this result has been first proven for $A_n$ and $D_n$ series in \cite{KS}. 
 In the next section we prove the relation (\ref{fxd}) (and hence (\ref{F=f}))  for the remaining simply-laced algebras, namely, for $E_n$, thus generalizing (\ref{zky}) to all 
 simply-laced simple Lie algebras. 

\section{ On validity of (\ref{fxd}) for all simply-laced algebras } 
In this section we prove the statement of the previous section, i.e. generalize the relation (\ref{fxd}) to all simply-laced algebras. 

The relation \ref{fxd} tells that at an integer $y$ the polynomial in $q=e^x$ in the denominator of $f(x,y)$ divides that in the numerator. 
In \cite{AM20PP}  it is shown that the necessary condition for it to take place is that the power of each factor in the denominator should divide the power of one of the
factors in the numerator. 
Indeed, this necessary condition is satisfied here. However, for a complete proof of \ref{fxd} one should consider each algebra separately. 

We claim that

\begin{eqnarray}\label{FXf}
	F_X(x,y)=f(x,y)
\end{eqnarray}
for any simply-laced  Lie algebra $X$. 

Take e.g. the $E_6$ algebra, for which the corresponding universal parameters in the minimal normalization are:
 $\alpha=-2, \beta=6, \gamma=8, t=12$.
We should calculate the sum  

\begin{eqnarray}\label{F}
F_{E_6} (x,y)= 6+	\sum_{m=0}^{y-1}	\sum_{\alpha_{+}} e^{x(y(\alpha,\rho) -m)}+e^{-x(y(\alpha,\rho) -m) }
\end{eqnarray}

First note the number of roots $n_L$ with a given height $L=(\alpha,\rho)$ among all roots. The set of couples $(L,n_L)$ with a non-zero $n_L$ is 

\begin{eqnarray}
 (-11,1),(-10,1),(-9,1),(-8,2),(-7,3),(-6,3),(-5,4),(-4,5),\\ \nonumber (-3,5),(-2,5),(-1,6),(0,6),(1,6),(2,5),(3,5),(4,5),(5,4),(6,3), \\ \nonumber
 (7,3),(8,2),(9,1),(10,1),(11,1)
\end{eqnarray}
which of course is symmetric w.r.t. the $L \leftrightarrow -L$. We also include the element $(0,6)$ in this list, which is just the first term $6$ in (\ref{F}).
Then, using this data, we note that the sum in (\ref{F}) is given by  

\begin{eqnarray}
F_{E_6}=	\phi(11y)+\phi(8y)+\phi(7y)+\phi(5y)+\phi(4y)+\phi(y) \\
	\phi(n)=\sum_{i=-n}^n q^i = \frac{q^{2n+1}-1}{q^n (q-1)} \\
	q=e^x
\end{eqnarray}

Combining the sums $\phi(11y)+\phi(8y)+\phi(5y)$ and $\phi(7y)+\phi(4y)+\phi(y)$, we get

\begin{eqnarray}
	\phi(11y)+\phi(8y)+\phi(5y)= \frac{(q^{9y}-1)(q^{5y+1}-q^{-11y})}{(q-1)(q^{3y}-1)} \\
	\phi(7y)+\phi(4y)+\phi(y) = \frac{(q^{9y}-1)(q^{y+1}-q^{-7y})}{(q-1)(q^{3y}-1)} \\
	F_{E_6}=\frac{(q^{9y}-1)}{(q-1)(q^{3y}-1)}(q^{4y}+1)(q^{y+1}-q^{-11y}) = \\
	\frac{(q^{9y}-1)(q^{8y}-1)(q^{y+1}-q^{-11y})}{(q-1)(q^{3y}-1)(q^{4y}-1)}
\end{eqnarray}
which can be easily checked to coincide with $f(x,y)$ for the universal parameters corresponding to $E_6$ algebra.

Literally similar calculations can be carried out for the remaining $E_7, E_8$ algebras, as well as for Krefl-Schwarz cases $A_n, D_n$, leading to the same conclusion.

\section {Universal-type representation of partition function for non-simply-laced algebras} \label{sect:univ}

Equations (\ref{fxd}), (\ref{F}) do not hold for non-simply-laced algebras. However, one can present the corresponding sum in a 
similar form, appropriate for further duality considerations \cite{M13,M21,KM}. The latter means that it can be presented as the ratio of a sum of exponents of 
$x$ (i.e. powers of $q=\exp x$) in the numerator and some sines in the denominator. So, we are aiming to represent $F_X$ as follows: 

\begin{eqnarray}\label{FX}
	F_X=    r+	\sum_{m=0}^{y-1}	\sum_{\alpha_{+}} \left( e^{x(y(\alpha,\rho) -m (\alpha,\alpha)/2)}+e^{-x(y(\alpha,\rho) -m (\alpha,\alpha)/2)}\right) 
	= \frac{A_X}{B_X}
\end{eqnarray}
where $X$ denotes an algebra of type $B,C,F$ or $G$, $r$ is its rank, $B_X$ is a product of a number of terms of the form $q^a-1$, and $A_X$ is a polynomial in $q$. 

One subtlety regarding the formulae (\ref{FX}), which makes them different from the (\ref{FXf}), is that in (\ref{FX}) one should explicitly mention the
 normalization of the scalar product. In (\ref{FXf}) both sides are invariant under the rescaling of the scalar product in the l.h.s.
  (with corresponding rescaling of the universal parameters in the r.h.s.), and the simultaneous appropriate rescaling of $x$. 
  However, in (\ref{FX}) a similar rescaling of the scalar product and $x$ leaves invariant only the l.h.s., whilst the ratio $A_X/B_X$ in the r.h.s is dependent only on $x$,
   thus changes under its rescaling. This means that when substituting the r.h.s. of (\ref{FX}) into the partition function \ref{Zintg} one should 
take the parameters $t$ and $\delta$ in the same normalization. The normalizations below are chosen to avoid the appearance of fractional powers of $q$. 

Now we present $F_X$ for all non-simply laced algebras. 

Let us consider the $B_n$ algebras. Normalization corresponds to $\alpha=-4$, i.e. the square of the long root is $4$.
 The corresponding representation we mentioned above is

\begin{eqnarray}\label{F2}
	F_{B_n}(x,y)= 
	\frac{A_{B_n}}{B_{B_n}} \\
	A_{B_n}=q^{4 n y+2}+q^{-4 (n-1) y}+   \\
	(q+1) \left(q^y-1\right) \left(q^{2 y}+1\right) \left(q^{2 n y}-1\right) \left(q^{y-2 n y}+q\right)-q^{4 y}-q^2 \\ 
	B_{B_n}=\left(q^2-1\right) \left(q^{4 y}-1\right),
\end{eqnarray}

For $C_n$ algebras we also choose the same normalization with square of long root 4. Then $F_X$ writes as

\begin{eqnarray}
	F_{C_n}= \frac{A_{C_n}}{B_{C_n}} \\
	B_{C_n}=(q^2-1)  \left(q^{2y}-1\right) \\
	A_{C_n}=	(q+1) q^y \left(q^{2 n y}-1\right) \left(q^{2 n y+1}-1\right)+\\
	\left(q^{2 y}-1\right) \left(q^{n y}-1\right) \left(q^{n y+1}-1\right) \left(q^{2 n y+1}-1\right)
\end{eqnarray}

For $F_4$, with the same normalization, we have 

\begin{eqnarray}
	F_{F_4}= \frac{A_{F_4}}{B_{F_4}} \\
	B_{F_4}=(q^2-1)   \\
	A_{F_4}=q^{-16 y} \left(q^{2 y}+1\right) \left(-q^{2 y}+q^{4 y}+1\right) \left(q^{12 y+1}-1\right) \times \\ \left(q^{5 y+1}-q^{8 y+1}+q^{9 y+1}+q^{14 y+1}+q^{5 y}-q^{6 y}+q^{9 y}+1\right)
\end{eqnarray}

For $G_2$ we use the normalization corresponding to the square of the long root to be equal to $6$. The corresponding $F_{G_2}$ function is

\begin{eqnarray}
	F_{G_2}= \frac{A_{G_2}}{B_{G_2}} \\
	B_{G_2}=q^3-1   \\
	A_{G_2}=q^{-9 y} \left(q^{6 y+1}-1\right) \times \\
	\left(q^{4 y+1}+q^{8 y+1}+q^{4 y+2}-q^{6 y+2}+q^{8 y+2}+q^{12 y+2}+q^{4 y}-q^{6 y}+q^{8 y}+1\right)
\end{eqnarray}

\section{Conclusion}

 By closing the contour of integration in (\ref{FintRep}) in the upper semiplane, one obtains the output by means of a sum of contributions of poles. 
As it was shown in a number of cases in \cite{M13,SGV,KM},  the contribution of so-called perturbative poles, 
i.e. those from $\sinh(x\delta)$, exactly coincides with the Gopakumar-Vafa partition function of corresponding dual topological string.
The corresponding contribution with the use of the newly derived expressions $F_X$ should be examined next, 
aiming at the interpretation of the initial partition function in terms of some (refined) topological strings, indeed if such strings exist.
 Obviously, that string would be the candidate for a dual description of the corresponding CS theory.
 
  We also hope that the exact closed expressions for partition functions of refined CS theory with arbitrary gauge group, derived in the present paper, 
  will find other applications, too.

\section{Acknowledgments.}

The work of MA was fulfilled within the Regional Doctoral Program on Theoretical and Experimental Particle Physics 
sponsored by VolkswagenStiftung.
The work of MA and RM is partially supported by the Science Committee of the Ministry of Science 
and Education of the Republic of Armenia under contracts  20AA-1C008, 21AG-1C060.

\begin{table}[ht] 
	\caption{Vogel's parameters}
	\begin{tabular}{|r|r|r|r|r|r|} 
		\hline & $\alpha$ &$\beta$  &$\gamma$  & $t=\alpha+\beta+\gamma$ & Line \\ 
		\hline  $su(N)$  & $-2$ & $2$ & $N$ & $N$ & $\alpha+\beta=0$ \\ 
		\hline $so(N)$ & $-2$  & $4$ & $N-4$ & $N-2$ & $ 2\alpha+\beta=0$ \\ 
		\hline  $ sp(N)$ & $-2$  & 1 & $N/2+2$ & $N/2+1$ & $ \alpha +2\beta=0$ \\ 
		\hline $Exc(a)$ & $-2$ & $a+4$  & $2a+4$ & $3a+6$ & $\gamma=2(\alpha+\beta)$\\ 
		\hline 
	\end{tabular}
	
	{For the exceptional 
		line $Exc(a)$ $a=-2/3,0,1,2,4,8$ for $G_2, SO(8), F_4, E_6, E_7,E_8 $, 
		respectively.} \label{tab:1}
\end{table}

\end{document}